\documentclass[12pt]{article}
\usepackage{amssymb}
\usepackage{amsmath}
\def\be{\begin{equation}}
\def\ee{\end{equation}}
\def\arr{\begin{array}{rll}}
\def\ea{\end{array}}
\def\bea{\begin{eqnarray}}
\def\eea{\end{eqnarray}}

\def\N2{$N{=}2$}

\def\>{\rangle}
\def\<{\langle}
\def\+{\dagger}
\def\={\ =\ }

\begin{document}
\renewcommand{\thefootnote}{\arabic{footnote}}
\begin{titlepage}
\setcounter{page}{0}
\begin{flushright}
$\quad$  \\
\end{flushright}
\vskip 3cm
\begin{center}
{\LARGE\bf Conformal symmetry in damped}
\vskip 0.5cm
{\LARGE\bf Pais-Uhlenbeck oscillator}\\
\vskip 2cm
$
\textrm{\Large Ivan Masterov}
$
\vskip 0.5cm
{\it
Tomsk State University of Control Systems and Radioelectronics\\
634050, Tomsk, Lenin Ave. 40, Russia}
\vskip 1cm
{E-mails: ivan.v.masterov@tusur.ru}

\end{center}
\vskip 1cm
\begin{abstract}
\noindent
Two Lagrangian formulations for describing of the damped harmonic oscillator have been introduced by Bateman. For these models we construct higher derivative generalization which enjoys the $l$-conformal Newton-Hooke symmetry. The dynamics of generalized systems corresponds to the damped Pais-Uhlenbeck oscillator for a particular choice of its frequencies.
\end{abstract}


\vskip 0.5cm


\end{titlepage}

\noindent
{\bf 1. Introduction}
\vskip 0.5cm
\noindent

As known, the Newton-Hooke ($\mathbb{NH}$) algebra \cite{Bacry} can be extended by conformal generators of dilatations and special conformal transformations in different ways. In general, such extensions are indexed by a positive integer of half-integer parameter $l$ \cite{Negro}. Motivated by the study of the nonrelativistic version of the AdS/CFT correspondence \cite{AdS_1,AdS_2}, the $l$-conformal $\mathbb{NH}$ algebra has recently attracted considerable attention.

The instance of $l=\frac{1}{2}$ has been the focus of most studies. For example, the harmonic oscillator \cite{Niederer} as well as its damped \cite{Martini}-\cite{DHO_2} and deformed \cite{KrLS,KN} versions exhibit invariance under the $l=\frac{1}{2}$-conformal $\mathbb{NH}$ group. Also, the same symmetry has been found in many-body conformal mechanics \cite{Calogero}-\cite{mb_2}.

For other values of $l$, dynamical realizations of the $l$-conformal $\mathbb{NH}$ group have been constructed in Refs. \cite{AV_1}-\cite{Baranovsky}. In particular, the analysis of the maximal kinematical invariance group of the so-called Pais-Uhlenbeck oscillator ($\mathbb{PU}$) \cite{PU}  shows that this system enjoys the $l$-conformal $\mathbb{NH}$ symmetry provided frequencies of oscillation form the arithmetic sequence of a special type \cite{AV_2} (see also \cite{AV_1,Masterov_1,Andrzejewski}).

The $\mathbb{PU}$ oscillator is higher derivative mechanical system. Higher derivative models generically show up instabilities in classical dynamics \cite{PU,Woodard}. Instances, which don't face this problem, are of interest as a physically viable. The $\mathbb{PU}$ oscillator is an example of such systems: for the case of distinct frequencies, the motion of this model is bounded and stable.

The stability of the $\mathbb{PU}$ oscillator may be lost in the presence of interaction\footnote{See, e.g., related discussion in Refs. \cite{Woodard,Pavsic_2}.}. An illustrative example of this situation for the one-dimensional fourth-order case has been considered in Ref. \cite{Nesterenko}. It was shown that an inclusion of the standard friction force, which leads to the damping of the harmonic oscillator, results in an appearance of exponentially growing solutions of the equation of motion for the $\mathbb{PU}$ oscillator.

The results in the work \cite{Nesterenko} have been generalized in Refs. \cite{Stephen,Pavsic}. It was demonstrated that there exists the force which tends to retard the motion of the fourth-order $\mathbb{PU}$ oscillator. In the presence of this force the system remains stable and oscillations are damped.

Lagrangian and Hamiltonian formulations as well as quantization of the damped fourth-order $\mathbb{PU}$ oscillator in one dimension have been considered in Ref. \cite{Gludko}. Various damping regimes for the same model have been analyzed in recent work \cite{Sanders}.  Any other aspects of the system remain completely unexplored. Moreover, to the best of our knowledge, the damped version of the $\mathbb{PU}$ oscillator for other orders and dimensions has not yet been considered in the literature. On the other hand, this issue is of interest because a switching on the oscillation's damping may preserve the classical stability of the system.

Taking into account conformal invariance of the damped harmonic oscillator, it can be expected that the same property may hold for the damped $\mathbb{PU}$ oscillator. The purpose of the present work is to demonstrate that the damped $\mathbb{PU}$ oscillator may exhibit the $l$-conformal $\mathbb{NH}$ symmetry. 

In the work \cite{Bateman} two Lagrangian formulations for the damped harmonic oscillator have been introduced. Our strategy is based on the construction of a higher derivative generalization of these systems which preserves the $l$-conformal $\mathbb{NH}$ symmetry. In particular, the first system, which is referred to as the Bateman-Caldirola-Kanai ($\mathbb{BCK}$) model, is generalized in the next section. In Sect. 3 we consider the second system, which is known as the Bateman model. In the concluding Sect. 4 we summarize our results and discuss further possible developments.

\vskip 0.5cm
\noindent
{\bf 2.  Higher derivative $\mathbb{BCK}$ model with the $l$-conformal $\mathbb{NH}$ symmetry}
\vskip 0.5cm

The equation of motion of the damped $d$-dimensional harmonic oscillator reads
\bea\label{dho}
&&
\left(\frac{d^2}{dt^2}+2\gamma\frac{d}{dt}+(\omega_0^2+\gamma^2)\right) x_i(t) = 0,
\eea
where $i=1,2,...,d$; $\gamma$ is the damping coefficient, which is positive throught the work; $\omega_0$ is the frequency of a free oscillation. The general solution of this equation is given by
\bea
&&
x_i(t) = e^{-\gamma t}\left(A_i \sin{(\omega_0 t)} + B_i \cos{(\omega_0 t)}\right),
\nonumber
\eea
where $A_i$ and $B_i$ are constants of integration.

The construction of Lagrangian and Hamiltonian formulations of the equation \eqref{dho} for $d=1$ has been initiated in the works \cite{Bateman}-\cite{Kanai}. In particular, in the work \cite{Bateman} Bateman introduced the action functional whose multidimensional version reads
\bea\label{CK}
&&
S = \frac{1}{2}\int dt\,e^{2\gamma t}x_i(t)\left(\frac{d^2}{dt^2}+2\gamma\frac{d}{dt}+(\omega_0^2+\gamma^2)\right)x_i(t).
\eea
While Caldirola \cite{Caldirola} and Kanai \cite{Kanai} have considered the Hamiltonian formulation which can be obtained from this action. By this reason the model \eqref{CK} is naturally refereed to as the Bateman-Caldirola-Kanai ($\mathbb{BCK}$) model.

As was established in the work \cite{JI}, the action \eqref{CK} can be obtained from the action functional of a free particle
\bea\label{fp}
&&
S = \frac{1}{2}\int dt\,x_i(t)\ddot{x}_i(t)
\eea
by applying Niederer-type transformation of the form
\bea\label{dho_nied}
&&
t\to\frac{1}{\omega_0}\tan{(\omega_0 t)}, \qquad x_i\to\frac{x_i e^{\gamma t}}{\cos{(\omega_0 t)}}.
\eea

To construct higher derivative counterpart of the $\mathbb{BCK}$ model \eqref{CK}, let us consider $(2l+1)$-order derivative generalization of the action \eqref{fp}, where $l$ is a positive integer or half-integer parameter. The action functional of the model reads \cite{Gomis}
\bea\label{HDFP}
&&
S = \frac{1}{2}\int dt\, \eta_{ij} x_ix_j^{(2l+1)}, \qquad \eta_{ij} = \left\{
\begin{aligned}
&
\delta_{ij} && \mbox{for half-integer }l && i,j=1,2,..,d,
\\[2pt]
&
\epsilon_{ij} && \mbox{for integer }l && i,j=1,2.
\end{aligned}
\right.
\eea
Let us note that this system for odd orders (i.e. for integer $l$) is defined only for $d=2$. 

At the next step, let us introduce the following generalization of the transformation \eqref{dho_nied}
\bea\label{HDNTck}
&&
t\to\frac{1}{\omega_0}\tan{(\omega_0 t)}, \quad x_i \to \frac{x_i e^{\gamma f(t)}}{\cos^{2l}{(\omega_0 t)}},
\eea
where $f(t)$ is now arbitrary differentiable function so that the quantity $\gamma f(t)$ being dimensionless. An applying these transformation to the action \eqref{HDFP} yields
\bea\label{HDCK}
&&
S = \frac{1}{2}\int dt \,e^{2\gamma f(t)}\eta_{ij} x_i(t) \mathcal{D}_{2l+1} x_j(t),
\eea
where
\bea\label{operator}
\mathcal{D}_{2l+1} = \left\{
\begin{aligned}
&
\prod_{k=0}^{l-1/2}\left(\frac{d^2}{dt^2}+2\gamma \dot{f}\frac{d}{dt} + ((2k+1)^2\omega_0^2 + \gamma^2\dot{f}^2 + \gamma\ddot{f})\right), &&  l=\frac{1}{2},\frac{3}{2},...,
\\[2pt]
&
\left(\frac{d}{dt}+\gamma\dot{f}(t)\right)\prod_{k=1}^{l}\left(\frac{d^2}{dt^2}+2\gamma \dot{f}\frac{d}{dt} + ((2k\omega_0)^2 + \gamma^2\dot{f}^2 + \gamma\ddot{f})\right), &&  l=1,2,...
\end{aligned}
\right.
\eea

The equation of motion of the model \eqref{HDCK} reads
\bea\label{DPU_eom}
&&
\mathcal{D}_{2l+1} x_i(t) = 0.
\eea
The general solution of this equation has the form
\bea
x_i(t) = \left\{
\begin{aligned}
&
e^{-\gamma f(t)}\sum_{k=0}^{l-1/2}\left(A_{k,i}\sin{((2k+1)\omega_0 t)} + B_{k,i}\cos{((2k+1)\omega_0 t)}\right), && \mbox{for half-integer }l,
\\[2pt]
&
e^{-\gamma f(t)}B_{0,i} + e^{-\gamma f(t)}\sum_{k=1}^{l}\left(A_{k,i}\sin{(2k\omega_0 t)} + B_{k,i}\cos{(2k\omega_0 t)}\right), && \mbox{for integer }l,
\end{aligned}
\right.
\nonumber
\eea
where $A_{k,i}$ and $B_{k,i}$ are constants of integration.

When $\gamma=0$, the model \eqref{HDCK} describes the $(2l+1)$-order $\mathbb{PU}$ oscillator whose frequencies of oscillation form the arithmetic sequence\footnote{Taking into account the form of the transformation \eqref{HDNTck}, we may conclude that the action \eqref{HDCK} may be also obtained from the model of the $\mathbb{PU}$ oscillator
\bea
&&
S = \int dt\, x_i(t) \mathcal{D}_{2l+1}\bigg|_{\gamma=0} x_i(t)
\nonumber
\eea
by applying the change $x_i(t) \to x_i(t) e^{\gamma f(t)}$.
}
\bea
&&
\begin{aligned}
&
\omega_0,\; 3\omega_0,\; 5\omega_0,\; ...,\; (2(l+1/2)-1)\omega_0 && \mbox{for half-integer } l,
\\[2pt]
&
2\omega_0,\; 4\omega_0,\; 6\omega_0,\; ..., 2l\omega_0 && \mbox{for integer }l. 
\end{aligned}
\nonumber
\eea
As known \cite{AV_1}-\cite{Andrzejewski}, such $\mathbb{PU}$ oscillator exhibits the $l$-conformal $\mathbb{NH}$ symmetry. The same is true for the case when $\gamma\neq 0$. Indeed, the coordinate transformation \eqref{HDNTck} allows us to obtain symmetries of the system \eqref{HDCK} from those of the model \eqref{HDFP} (for a more details see Appendix). So, the model \eqref{HDCK} is invariant under transformations whose generators read
\bea\label{hdCK_gen}
\begin{aligned}
&
H = \frac{\partial}{\partial t} - \gamma\dot{f}(t) x_i\frac{\partial}{\partial x_i}, && D = \frac{\sin{(2\omega_0 t)}}{2\omega_0} H + l\cos{(2\omega_0 t)} x_i\frac{\partial}{\partial x_i},
\\[2pt]
&
M_{ij} = x_i\frac{\partial}{\partial x_j} - x_j \frac{\partial}{\partial x_i}, && K = \frac{\sin^{2}{(\omega_0 t)}}{\omega_0^{2}} H + \frac{l}{\omega_0}\sin{(2\omega_0 t)} x_i \frac{\partial}{\partial x_i},
\\[2pt]
&
 C_i^{(n)} = \frac{e^{-\gamma f(t)}}{\omega_0^{n}}\sin^{n}{(\omega_0 t)} \cos^{2l-n}{(\omega_0 t)} \frac{\partial}{\partial x_i}, && n = 0,1,..,2l.
\end{aligned}
\eea

Let us note that the Lagrangian of the model \eqref{HDCK} explicitly depends on time if $f(t)\neq const$. So, this system is not invariant under time translations. Instead of this, higher-derivative $\mathbb{BCK}$ model exhibits an invariance under generalized time translations which are generated by the operator $H$.

It can be straightforwardly verified that non-vanishing commutation relations between generators \eqref{hdCK_gen} read
\bea
\begin{aligned}
&
[H,D] = H -2\omega_0^2 K,  && [H,C_i^{(n)}] = nC_i^{(n-1)} + (n-2l)\omega_0^2 C_i^{(n+1)},
\\[2pt]
&
[H,K] = 2D, && [D,C_i^{(n)}] = (n-l)C_i^{(n)},
\\[2pt]
&
[D,K] = K, && [K,C_i^{(n)}] = (n-2l)C_i^{(n+1)},
\\[2pt]
&
[M_{ij},C_k^{(n)}] = -\delta_{ik}C_j^{(n)} + \delta_{jk}C_i^{(n)} && [M_{ij},M_{kp}] = -\delta_{ik}M_{jp} - \delta_{jp}M_{ik} + \delta_{ip}M_{jk} + \delta_{jk}M_{ip}.
\end{aligned}
\eea
This is the $l$-conformal $\mathbb{NH}$ algebra \cite{Negro,AV_3}. The generators $H$, $D$, and $K$ form $so(1,2)$ subalgebra.

The oscillations of the $\mathbb{PU}$ oscillator are damped if $f(t)$ is a positive-definite increasing function. So, in this case the model \eqref{HDCK} and the equation \eqref{DPU_eom} correspond to the damped $\mathbb{PU}$ oscillator which enjoys the $l$-conformal $\mathbb{NH}$ symmetry.

\vskip 0.5cm
\noindent
{\bf 3. Higher derivative Bateman model with the $l$-conformal $\mathbb{NH}$ symmetry}
\vskip 0.5cm
In the same work \cite{Bateman}, Bateman has introduced one more Lagrangian formulation of the equation \eqref{dho} for $d=1$. An analogue of this model for the case of arbitrary $d$ can be written in the following form
\bea\label{BM}
&&
S = \int dt\,y_i(t) \left(\frac{d^2}{dt^2}+2\gamma\frac{d}{dt}+(\omega_0^2+\gamma^2)\right)x_i(t).
\eea
In the literature this system is refereed to as the Bateman model.

In contrast to \eqref{CK}, the Lagrangian in the model \eqref{BM} does not depend explicitly on time, but new dynamical variable $y_i(t)$ is appeared. The dynamics of the Bateman model is described by \eqref{dho} and by the equation of the form
\bea\label{aho}
&&
\ddot{y}_i(t) - 2\gamma \dot{y}_i(t) + (\omega_0^2+\gamma^2) y_i(t) = 0.
\eea
The general solution of this equation is given by
\bea
&&
y_i(t) = e^{\gamma t}\left(\tilde{A}_i \sin{(\omega_0 t)} + \tilde{B}_i \cos{(\omega_0 t)}\right),
\nonumber
\eea
where $\tilde{A}_i$ and $\tilde{B}_i$ are constants of integration. So, oscillations of the particle with coordinates $y_i(t)$ are amplified. As a consequence, this particle is often called as the amplified harmonic oscillator.

The Bateman model is invariant under time translations and consequently the total energy of the system is conserved. This means that the energy, which is dissipated by the damped harmonic oscillator, simultaneously is absorbed by the amplifed harmonic oscillator.

Let us note that the model \eqref{BM} can be obtained from the action functional
\bea\label{fbm}
&&
S = \int dt \,y_i(t) \ddot{x}_i(t)
\eea
by applying the following Niederer-type transformation
\bea\label{HDNT_1}
&&
t\to\frac{1}{\omega_0}\tan{(\omega_0 t)}, \qquad x_i(t) \to \frac{x_i(t)e^{\gamma t}}{\cos{(\omega_0 t)}}, \qquad y_i(t) \to \frac{y_i(t)e^{-\gamma t}}{\cos{(\omega_0 t)}}.
\eea

By analogy with the analysis in the previous section, let us consider the following higher derivative analogue of the action \eqref{fbm}
\bea\label{hfbm}
&&
S = \int dt\, y_i(t) x_i^{(2l+1)}(t)
\eea
and the Niederer-type transformation of the form
\bea\label{HDNTb}
&&
t\to\frac{1}{\omega_0}\tan{(\omega_0 t)}, \quad x_i(t) \to \frac{x_i(t)e^{\gamma f(t)}}{\cos^{2l}{(\omega_0 t)}}, \quad y_i(t) \to \frac{y_i(t)e^{-\gamma f(t)}}{\cos^{2l}{(\omega_0 t)}}.
\eea
An applying \eqref{HDNTb} to \eqref{hfbm} yields
\bea\label{HDB}
&&
S = \int dt\, y_i(t) \mathcal{D}_{2l+1} x_i(t),
\eea
where the operator $\mathcal{D}_{2l+1}$ is defined in \eqref{operator}. In contrast to \eqref{HDCK}, for integer $l$ the model \eqref{HDB} is suitable for arbitrary $d$.

The dynamics of the particle, which parametrized by $x_i(t)$, is governed by the equation \eqref{DPU_eom}. While dynamical variable $y_i(t)$ is satisfied to
\bea
&&
\overline{\mathcal{D}}_{2l+1} y_i(t) = 0,
\nonumber
\eea
where the operator $\overline{\mathcal{D}}_{2l+1}$ may be obtained from $\mathcal{D}_{2l+1}$ by applying the time reversion $t\to -t$. So, if $f(t)$ is a positive-definite increasing function, $x_i(t)$ and $y_i(t)$ correspond to the damped and amplified $\mathbb{PU}$ oscillators, respectively.

The symmetry transformations of the higher derivative Bateman model \eqref{HDB} can be derived from those of the model \eqref{hfbm} with the aid of the link \eqref{HDNTb} (the details are omitted to Appendix). The corresponding generators read
\bea
\begin{aligned}
&
H = \frac{\partial}{\partial t} -\gamma\dot{f}x_i\frac{\partial}{\partial x_i} + \gamma\dot{f} y_i\frac{\partial}{\partial y_i}, && D = \frac{\sin{(2\omega_0 t)}}{2\omega_0}H + l\cos{(2\omega_0 t)}\left(x_i\frac{\partial}{\partial x_i} + y_i\frac{\partial}{\partial y_i}\right),
\\[2pt]
&
G_{ij} = x_i\frac{\partial}{\partial x_j} - y_j\frac{\partial}{\partial y_i}, && K = \frac{\sin^{2}{(\omega_0 t)}}{\omega_0^2} H + \frac{l}{\omega_0}\sin{(2\omega_0 t)}\left(x_i\frac{\partial}{\partial x_i} + y_i\frac{\partial}{\partial y_i}\right),
\\[2pt]
&
C_{1,i}^{(n)} = \frac{e^{\gamma f(t)}}{\omega_0^{n}}\sin^{n}{(\omega_0 t)}\cos^{2l-n}{(\omega_0 t)} \frac{\partial}{\partial y_i}, && C_{0,i}^{(n)} = \frac{e^{-\gamma f(t)}}{\omega_0^{n}}\sin^{n}{(\omega_0 t)}\cos^{2l-n}{(\omega_0 t)} \frac{\partial}{\partial x_i}.
\end{aligned}
\nonumber
\eea

These generators obey the following non-vanishing commutation relations:
\bea\label{alg}
&&
\begin{aligned}
&
[H,D] = H -2\omega_0^2 K, && [H,C_{\alpha,i}^{(n)}] = nC_{\alpha,i}^{(n-1)} + (n-2l)\omega_0^2 C_{\alpha,i}^{(n+1)},
\\[2pt]
&
[H,K] = 2D, && [D,C_{\alpha,i}^{(n)}] = (n-l)C_{\alpha,i}^{(n)},
\\[2pt]
&
[D,K] = K, && [K,C_{\alpha,i}^{(n)}] = (n-2l)C_{\alpha,i}^{(n+1)}, && 
\\[2pt]
&
[G_{ij},G_{kl}] = \delta_{jk} G_{il} - \delta_{il}G_{kj}, && [G_{ij},C_{\alpha,k}^{(r)}] = \delta_{\alpha,1}\delta_{kj}C_{1,i}^{(n)} - \delta_{\alpha,0}\delta_{ik}C_{0,j}^{(n)}.
\end{aligned}
\eea

A few comments are in order. At first, the algebra \eqref{alg} contains abelian ideal $V^{2d\times(2l+1)}$ which is spanned by vector generators $C_{\alpha,i}^{(n)}$. The generators $H$, $D$, and $K$ form $so(1,2)$ subalgebra while the generators $G_{ij}$ correspond to $gl(d,\mathbb{R})$ subalgebra. So, the algebra \eqref{alg} has the following semi-direct sum structure $(so(1,2)\oplus gl(d,\mathbb{R}))\ltimes V^{2d(2l+1)}$.

At second, two sets of generators $(H,D,K,C_{0,i}^{(n)},M_{ij})$ and $(H,D,K,C_{1,i}^{(n)},M_{ij})$, where $M_{ij} \equiv G_{ij}-G_{ji}$, both obey the structure relations of the $l$-conformal $\mathbb{NH}$ algebra.

At third, if $f(t)=t$ the Lagrangian of the model \eqref{HDB} does not explicitly depend on time. So, in this case the system is invariant under time translations. The corresponding generator is appeared as the linear combination $H+G_{ii}$.

\vskip 0.5cm
\noindent
{\bf 4. Conclusion}
\vskip 0.5cm

To summarize, in this work we have constructed a higher derivative generalization of the $\mathbb{BCK}$ and Bateman models. The first generalized system describes the damped $\mathbb{PU}$ oscillator and enjoys the $l$-conformal $\mathbb{NH}$ symmetry. The second generalized system involves the damped and the amplified $\mathbb{PU}$ oscillators. The symmetry algebra of the model is more wide than the $l$-conformal $\mathbb{NH}$ algebra and has the structure $(so(1,2)\oplus gl(d,\mathbb{R}))\ltimes V^{2d(2l+1)}$.

Some words about further possible developments. In the present work we were interested in the damped $\mathbb{PU}$ oscillator which exhibits the $l$-conformal $\mathbb{NH}$ symmetry. In this case all frequencies of oscillation are proportional to one quantity $\omega_0$ while the damping process is characterized by one constant $\gamma$ and by one function $f(t)$. But the damped $\mathbb{PU}$ oscillator as a stable higher derivative mechanical system may be of interest without reference to any symmetry aspects. This may motivate to investigate the damped $\mathbb{PU}$ oscillator of a more general type: for arbitrary frequencies of oscillation and for more general damped regime. For example, a more general form of a damping for the fourth-order $\mathbb{PU}$ oscillator corresponds to the following evolution
\bea
&&
x_i(t) = e^{-\gamma_0 f_0(t)}(A_{0,i} \sin{(\omega_0 t)} + B_{0,i} \cos{(\omega_0 t)}) + e^{-\gamma_1 f_1(t)}(A_{1,i} \sin{(\omega_1 t)} + B_{1,i} \cos{(\omega_1 t)}),
\nonumber
\eea
where $A_{\alpha,i}$, $B_{\alpha,i}$, $\omega_\alpha$, $\gamma_\alpha$ with $\alpha=0,1$ are constants while $f_{\alpha}(t)$ are positive-definite increasing functions.

As known, the $\mathbb{PU}$ oscillator is milti-Hamiltonian system, i.e. admits a variety of Hamiltonian formulations \cite{Kosinski}-\cite{Masterov_3}. It would be interesting to explore possible Hamiltonian structures for the damped $\mathbb{PU}$ oscillator.

Taking into account that the models \eqref{HDCK}, \eqref{HDB} are not invariant under pure time translations, it is of interest to adapt standard supersymmetry transformations in such a way that to obtain supersymmetric generalizations of the models derived in the present work.

These issues will be studied elsewhere.

\vskip 0.5cm
\noindent
{\bf Appendix. About symmetries of the models \eqref{HDCK}, \eqref{HDB}}
\vskip 0.5cm
The free particle model \eqref{HDFP} is invariant under transformations which are generated by the following operators \cite{Gomis} (see also \cite{Horvathy_2}):
\bea\label{gens}
\begin{aligned}
&
H = \frac{\partial}{\partial t}, \; D = t\frac{\partial}{\partial t} + lx_i\frac{\partial}{\partial x_i}, \; K = t^2\frac{\partial}{\partial t} + 2ltx_i\frac{\partial}{\partial x_i},
\\[2pt]
&
M_{ij} = x_i\frac{\partial}{\partial x_j} - x_j\frac{\partial}{\partial x_i}, \; C_i^{(n)} = t^{n}\frac{\partial}{\partial x_i}.
\end{aligned}
\eea
The Noether integrals of motion which correspond to these symmetries read
\bea
&&
\mathcal{H} = \sum_{n=0}^{2l-1}(-1)^{n}x_i^{(n+1)}x_i^{(2l-n)}, \quad \mathcal{D} = t\mathcal{H} - \sum_{n=0}^{2l}(-1)^{n}(l-n) x_i^{(n)} x_i^{(2l-n)},
\nonumber
\\[2pt]
&&\label{consts}
\mathcal{K} = -t^2\mathcal{H} + 2t\mathcal{D} - \sum_{n=1}^{2l}(-1)^{n}n(2l-n+1)x_i^{(n-1)}x_i^{(2l-n)},
\\[2pt]
&&
\mathcal{M}_{ij} = \sum_{n=0}^{2l}(-1)^{n}x_i^{(n)}x_j^{(2l-n)}, \quad \mathcal{C}_i^{(n)} = \sum_{k=0}^{n}\frac{(-1)^{k} n!}{(n-k)!} t^{n-k} x_i^{(2l-k)}.
\nonumber
\eea
We denote the conserved charge by the same letters as the corresponding symmetry generator but in calligraphic style.

The generators of symmetry transformations \eqref{hdCK_gen} and the integrals of motion for the model \eqref{HDCK} can be derived from \eqref{gens} and \eqref{consts} by applying the transformation \eqref{HDNTck} and by implementation of the linear change
\bea\label{redef}
&&
H \to H + \omega_0^2 K, \qquad \mathcal{H} \to \mathcal{H} + \omega_0^2 \mathcal{K}.
\eea

The symmetries of the model \eqref{fbm} correspond to the generators of the form
\bea\label{genb}
&&
\begin{aligned}
&
H = \frac{\partial}{\partial t}, \quad \; G_{ij} = x_i\frac{\partial}{\partial x_j} - y_j\frac{\partial}{\partial y_i}, && C_{0,i}^{(n)} = t^{n}\frac{\partial}{\partial x_i}, \quad C_{1,i}^{(n)} = t^n\frac{\partial}{\partial y_i},
\\[2pt]
&
D = t\frac{\partial}{\partial t} + l x_i\frac{\partial}{\partial x_i} + l y_i\frac{\partial}{\partial y_i}, && K = t^2\frac{\partial}{\partial t} + 2ltx_i\frac{\partial}{\partial x_i} + 2lty_i\frac{\partial}{\partial y_i}.
\end{aligned}
\eea
The constants of motion associated with these symmetries read
\bea\label{constb}
\begin{aligned}
&
\mathcal{H} = \sum_{n=0}^{2l-1}(-1)^{n}x_i^{(n+1)}y_i^{(2l-n)}, \qquad \mathcal{D} = t\mathcal{H} - \sum_{n=0}^{2l}(-1)^{n}(l-n) x_i^{(n)} y_i^{(2l-n)},
\\[2pt]
&
\mathcal{K} = -t^2\mathcal{H} + 2t\mathcal{D} - \sum_{n=1}^{2l}(-1)^{n}n(2l-n+1)x_i^{(n-1)}y_i^{(2l-n)}, \; \mathcal{G}_{ij} = \sum_{n=0}^{2l}(-1)^{n}x_i^{(n)}y_j^{(2l-n)},
\\[2pt]
&
\mathcal{C}_{0,i}^{(n)} = \sum_{k=0}^{n}\frac{(-1)^{k}n!}{(n-k)!} t^{n-k} y_i^{(2l-k)}, \quad \mathcal{C}_{1,i}^{(n)} = \sum_{k=0}^{n}\frac{(-1)^{k} n!}{(n-k)!} t^{n-k} x_i^{(2l-k)}.
\end{aligned}
\eea
While applying the transformation \eqref{HDNTb} to \eqref{genb}, \eqref{constb} and further change \eqref{redef} yield the corresponding expressions for the model \eqref{HDB}.

\fontsize{10}{13}\selectfont


\begin{thebibliography}{nn}
\bibitem{Bacry}
H. Bacry, J.M. L\'{e}vy-Leblond, {\it Possible kinematics}, J. Math. Phys. {\bf 9} (1968) 1605.
\bibitem{Negro}
J. Negro, M.A. del Olmo, A. Rodriguez-Marco, {\it Nonrelativistic conformal groups}, J. Math. Phys. {\bf 38} (1997) 3786.
\bibitem{AdS_1}
D.T. Son, {\it Toward an AdS/cold atoms correspondence: a geometric realization of the Schrodinger symmetry}, Phys. Rev. D {\bf 78} (2008) 046003, arXiv:0804.3972[hep-th].
\bibitem{AdS_2}
K. Balasubramanian, J. McGreevy, {\it Gravity duals for non-relativistic CFTs}, Phys. Rev. Lett. {\bf 101} (2008) 061601, arXiv:0804.4053[hep-th].
\bibitem{Niederer}
U. Niederer, {\it The maximal kinematical invariance group of the harmonic oscillator}, Helv. Phys. Acta {\bf 46} (1973) 191.
\bibitem{Martini}
R. Martini, P.H.M. Kersten, {\it Contact symmetries of general linear second-order ordinary differential equations}, J. Phys. A {\bf 16} (1983) L455.
\bibitem{DHO_1}
J.M. Cerver\'{o}, J. Villarroel, {\it SL(3,R) realisations and the damped harmonic oscillator}, J. Phys. A {\bf 17} (1984) 1777.
\bibitem{DHO_2}
M. Cariglia, C. Duval, G.W. Gibbons, P.A. Horvathy, {\it Eisenhart lifts and symmetries of time-dependent systems}, Ann. Phys. {\bf 373} (2016) 631, arXiv:1605.01932[hep-th].
\bibitem{KrLS}
S. Krivonos, O. Lechtenfeld, A. Sorin, {\it Minimal realization of l-conformal Galilei algebra, Pais-Uhlenbeck oscillators and their deformation}, JHEP {\bf 10} (2016) 078, arXiv:1607.03756[hep-th].
\bibitem{KN}
S. Krivonos, A. Nersessian, {\it SU(1,2) invariance in two-dimensional oscillator}, JHEP {\bf 02} (2017) 006, arXiv:1610.02499[hep-th].
\bibitem{Calogero}
F. Calogero, C. Marchioro, {\it Exact bound states of some n-body systems with two- and three-body forces}, J. Math. Phys. {\bf 14} (1973) 182.
\bibitem{Wolfes}
J. Wolfes, {\it On the three-body linear problem with three-body interaction}, J. Math. Phys. {\bf 15} (1974) 1420.
\bibitem{mb_1}
A. Galajinsky, {\it N=2 superconformal Newton-Hooke algebra and many-body mechanics}, Phys. Lett. B {\bf 680} (2009) 510, arXiv:0906.5509[hep-th].
\bibitem{mb_2}
A. Galajinsky, {\it Conformal mechanics in Newton-Hooke spacetime}, Nucl. Phys. B {\bf 832} (2010) 586, arXiv:1002.2290[hep-th].
\bibitem{AV_1}
A. Galajinsky, I. Masterov, {\it Dynamical realizations of $l$-conformal Newton-Hooke group}, Phys. Lett. B {\bf 723} (2013) 190, arXiv:1303.3419[hep-th].
\bibitem{AV_2}
K. Andrzejewski, A. Galajinsky, J. Gonera, I. Masterov, {\it Conformal Newton-Hooke symmetry of Pais-Uhlenbeck oscillator}, Nucl. Phys. B {\bf 885} (2014) 150, arXiv:1402.1297[hep-th].
\bibitem{Masterov_1}
I. Masterov, {\it Dynamical realizations of $\,\mathcal{N}=1$ $l$-conformal Galilei superalgebra}, J. Math. Phys. {\bf 55} (2014) 102901, arXiv:1407.1438[hep-th].
\bibitem{Andrzejewski}
K. Andrzejewski, {\it Conformal Newton-Hooke algebras, Niederer's transformation and Pais-Uhlenbeck oscillator}, Phys. Lett. B {\bf 738} (2014) 405, arXiv:1409.3926[hep-th].
\bibitem{Baranovsky}
O. Baranovsky, {\it Higher-derivative generalization of conformal mechanics}, J. Math. Phys. {\bf 58} (2017) 082903, arXiv:1704.04880[hep-th].
\bibitem{PU}
A. Pais, G.E. Uhlenbeck, {\it On field theories with non-localized action}, Phys. Rev. {\bf 79} (1950) 145.
\bibitem{Woodard}
R. Woodard, {\it Avoiding dark energy with 1/R modifications of gravity}, Lect. Notes Phys. {\bf 720} (2007) 403, astro-ph/0601672.
\bibitem{Pavsic_2}
M. Pav\v{s}i\v{c}, {\it On negative energies, strings, branes, and braneworlds: A review of novel approaches}, Int. Mod. Phys. A {\bf 35} (2020) 2030020, arXiv:2012.04976[hep-th].
\bibitem{Nesterenko}
V.V. Nesterenko, {\it Instability of classical dynamics in theories with higher derivatives}, Phys. Rev. D {\bf 75} (2007) 087703, hep-th/0612265.
\bibitem{Stephen}
N.G. Stephen, {\it On the Ostrogradski instability for higher-order derivative theories and a pseudo-mechanical energy}, J. Sound and Vibration {\bf 310} (2008) 729.
\bibitem{Pavsic}
M. Pav\v{s}i\v{c}, {\it Pais-Uhlenbeck oscillator with benign friction force}, Phys. Rev. D {\bf 87} (2013) 107502, arXiv:1304.1325[gr-qc].
\bibitem{Gludko}
N. Glud'ko, {\it Hamiltonian formalism and stability of dissipative dynamical systems with higher derivatives}, Master's thesis (2017) (https://vital.lib.tsu.ru/vital/access/manager/Repository/vital:4412, in Russian).
\bibitem{Sanders}
J.W. Sanders, {\it Fourth-order dynamics of the damped harmonic oscillator}, \\ arXiv:2109.06034[physics.class-ph].
\bibitem{Bateman}
H. Bateman, {\it On dissipative systems and related variational principles}, Phys. Rev. {\bf 38} (1931) 815.
\bibitem{Caldirola}
P. Caldirola, {\it Forze non conservative nell meccanica quantistica}, Nuovo Cim. {\bf 18} (1941) 393.
\bibitem{Kanai}
E. Kanai, {\it On the quantization of the dissipative systems}, Prog. Theor. Phys. {\bf 3} (1948) 440.
\bibitem{JI}
G. Junker, A. Inomata, {\it Transformation of the free propagator to the quadratic propagator}, Phys. Lett. A {\bf 110} (1985) 195.
\bibitem{AV_3}
A. Galajinsky, I. Masterov, {\it Remarks on l-conformal extension of the Newton-Hooke algebra}, Phys. Lett. B {\bf 702} (2011) 265, arXiv:1104.5115[hep-th].
\bibitem{Kosinski}
K. Bolonek, P. Kosinski, {\it Hamiltonian structures for Pais-Uhlenbeck oscillator}, Acta Phys. Polon. B {\bf 36} (2005) 2115, quant-ph/0501024.
\bibitem{KLS}
D.S. Kaparulin, S.L. Lyakhovich, A.A. Sharapov, {\it Classical and quantum stability of higher-derivative dynamics}, Eur. Phys. J. C {\bf 74} (2014) 3072, arXiv:1407.8481[hep-th].
\bibitem{KL}
D.S. Kaparulin, S.L. Lyakhovich, {\it On the stability of nonlinear oscillator with higher derivatives}, Russ. Phys. J. {\bf 57} (2015) 1561.
\bibitem{Masterov_2}
I. Masterov, {\it An alternative Hamiltonian formulation for the Pais-Uhlenbeck oscillator}, Nucl. Phys. B {\bf 902} (2016) 95, arXiv:1505.02583[hep-th].
\bibitem{Masterov_3}
I. Masterov, {\it The odd-order Pais-Uhlenbeck oscillator}, Nucl. Phys. B {\bf 907} (2016) 495, arXiv:1603.07727[math-ph].
\bibitem{Gomis}
J. Gomis, K. Kamimura, {\it Schrodinger equations for higher order non-relativistic particles and N-Galilean conformal symmetry}, Phys. Rev. D {\bf 85} (2012) 045023, arXiv:1109.3773[hep-th].
\bibitem{Horvathy_2}
C. Duval, P. Horvathy, {\it Conformal Galilei groups, Veronese curves, and Newton-Hooke spacetimes}, J. Phys. A {\bf 44} (2011) 335203, arXiv:1104.1502[hep-th].
\end{thebibliography}
\end{document}